\documentclass{article}
\usepackage{amsmath}
\usepackage{amsfonts}
\usepackage{graphicx}
\usepackage{url}
\usepackage{authblk}
\usepackage{enumitem}

\newcommand{\mat}[1]{\mathbf{#1}}

\begin{document}

\title{Forecasting the impact of state pension reforms in post-Brexit England and Wales using microsimulation and deep learning}
\author{Agnieszka Werpachowska\footnote{aw@averisera.uk}}
\affil{Averisera Ltd\\ London, UK}

\maketitle

\begin{abstract}
We employ stochastic dynamic microsimulations to analyse and forecast the pension cost dependency ratio for England and Wales from 1991 to 2061, evaluating the impact of the ongoing state pension reforms and changes in international migration patterns under different Brexit scenarios. To fully account for the recently observed volatility in life expectancies, we propose mortality rate model based on deep learning techniques, which discovers complex patterns in data and extrapolated trends. Our results show that the recent reforms can effectively stave off the ``pension crisis'' and bring back the system on a sounder fiscal footing. At the same time, increasingly more workers can expect to spend greater share of their lifespan in retirement, despite the eligibility age rises. The population ageing due to the observed postponement of death until senectitude often occurs with the compression of morbidity, and thus will not, perforce, intrinsically strain healthcare costs. To a lesser degree, the future pension cost dependency ratio will depend on the post-Brexit relations between the UK and the EU, with ``soft'' alignment on the free movement lowering the relative cost of the pension system compared to the ``hard'' one. In the long term, however, the ratio has a rising tendency.
\end{abstract}


\section{Introduction}
\label{sec:intro}

The social and economic implications of a rapidly ageing population, a consequence of low fertility rates and rising longevity, are becoming increasingly apparent in the UK and other developed countries. A shrinking share of workers has to fill the labour void left by the retiring ones and concurrently provide for their support, as reflected in the growing old-age dependency ratio~\cite{Werpachowska}. To reduce the strain the new demographic situation places on public services, in particular the state pensions, benefits, and health care, the UK introduced a broad reform of the state pension system, raising the retirement age for all born after 1950. The former pension ages of 65 for men and 60 for women introduced in 1948 are due to be equalised by 2018, and subsequently increase for everybody to 68 by 2046. The economic and demographic effects of approaching Brexit will add to the magnitude of population changes against which the affordability of the pension system will be tested in the following decades.

This work evaluates the impact of the state pension reforms on the pension cost dependency ratio (namely, the number of people at or above the retirement age to people between ages 15 and that retirement age) under different post-Brexit scenarios. To this end, we perform stochastic dynamic microsimulations of the England and Wales (E\&W) population based on available historical data, as described in Ref.\,\cite{Werpachowska}, from 1991 (and forecasting beyond 2014) to 2061. The simulated future scenarios focus on changing migration patterns between the UK and the EU after the Brexit. In each scenario, we compare the dependency ratios assuming the state pension age with and without the past reforms. To fully account for the increasing trends in life expectancies, we have implemented a custom extrapolation model for mortality rates, based on deep learning techniques. The historical overview of state pension reforms and demographic changes in the UK in Sec.\,\ref{sec:background} provides insight in the presented analysis and results. Section\,\ref{sec:deeplearning} describes the modelling framework for neural networks and the obtained mortality rate forecasts. The microsimulation results for the pension cost dependency ratio are presented in Sec.\,\ref{sec:microsimulation} followed by the summary.

\section{Historical background and future scenarios}
\label{sec:background}

The foundations of the British social security system have been laid at the early stages of the state formation. A host of demographic, economic, political, and religious changes in the medieval England impelled the so-called Poor Law~\cite{macnicol}. For several centuries, it formed the basis of the government provision of relief for the ``impotent poor'' while forcing able-bodied ``vagrants'' into the infamous workhouses. Yet the idea of retirement and state pensions dawned with the 18th century industrial revolution~\cite{maseres}. At its early labour-intensive stage, millions of workers migrated to cities, to be eventually forced out from their factory jobs by rapid technological developments and the inflow of younger competition from the declining agriculture. Simultaneously, demographic trends characteristic of a maturing economy conduced to the growing share of ``dependants'' at the top of the population pyramid, boosting the case for the old age pension. Last but not least, industrialisation created a wealth to fund it.

The campaign for social welfare reform culminated in the Old-Age Pensions Act 1908 which established a gender-blind, non-contributory means-tested state pension for persons over the age of 70. From the start, the policy has been shaped by a multitude of conflicting agendas, from the conservative desire to limit income redistribution, through the Treasury policy of reducing government spending, to socialist demands of a tax-funded and universal subsistence scheme. Less visible (at least until the post-WW1 period) but as important as the class conflict was the gender division: although women constituted the majority of the aged poor~\cite{Caradog}, the debate over pensions focused on the plight of ``worn-out'' male worker. As a result, women's pension rights were eroded in post-1908 reforms.

Meeting the demands of emerging interest groups for increasing the level and scope of the pensions was considered ``fiscally unsound'' in the atmosphere of austerity prevalent in the 1920s~\cite{Blaikie}. Thus, the Conservative Party and civil service mandarins pushed for the switch to a contributory system (funded by worker's contributions paid during lifetime and available only to those with a minimum employment record). Such a system was expected to be more resistant to demands for increasing the pension amount and to require much less redistribution of income between the wealthy and the lower classes. This reactionary policy met with weak opposition from the Labour parliamentary leadership and culminated in the Widows, Orphans and Old Age Contributory Pensions Act 1925. As well as introducing a non-means tested, contributory pension system, the policy lowered the pension age from 70 to 65 for both men and women in hope to alleviate the unemployment of younger workers. Since many women did not have paid employment, they gained the right to a full pension based on their husband's contributions, which tied them financially to marriage and created a special group of interest, National Spinsters' Association. In 1940, female pension age was reduced to 60 and supplementary means-tested pensions introduced to address the problem of poverty among older unmarried women~\cite{Ginn}.

In 1941, UK government asked Sir William Beveridge to produce a comprehensive report, proposing a new ``from the cradle to the grave'' social welfare system for post-war Britain. Resulting from his work, the National Insurance Act 1946 
introduced a universal flat-rate contributory pension higher than the pre-war amount: \pounds 1 6s for a single person and \pounds 2 2s for a married couple, funded from workers' National Insurance contributions and payable from the age of 65 for men and 60 for women. National Insurance contributions were mandatory for everyone except married women, and no benefits were provided for divorced women. Thus, the new system reinforced the dependency of women on their husbands' pensions, a problem which became increasingly apparent with the post-WW2 changes in lifestyle (higher prevalence of single parenthood and divorce). Another important feature of the National Insurance was the retirement condition for the payment of pension (the aim of which was to prevent pensioners from undercutting younger workers in the labour market). However, because of post-war labour shortage, old men were encouraged to remain working past the pension eligibility age and two-thirds of them chose to do so~\cite{NA}.

To satisfy the needs of higher-earning workers, the National Insurance Act 1959 (implemented in 1961) introduced an additional earnings-based top-up pension, replaced in 1978 by the State Earnings-related State Pension Scheme. Workers who decided to provision privately for an additional pension could choose to opt out of SERPS and pay lower NI contributions. SERPS was replaced in 2002 by the State Second Pension Scheme, with the aim to skew the benefits of additional pensions in favour of low and moderate earners at the expense of the wealthier workers, and improve the situation of carers and disabled persons. The optional additional pensions have been abolished in 2016, when they have been all replaced by a single flat-rate state pension, bringing the UK pension system back to its Beveridgean roots and ending its post-war foray into the Bismarckian regime.\footnote{European state pension systems range from a Bismarckian design with no redistribution and pensions that are earnings-related, to a Beveridgean one with flat pensions. The latter comprises the main part of the UK system of public support for pensioners, next to earnings-related benefits, flat-rate non-contributory benefits and means-tested benefits. It is worth noting that although its design translates into replacement rates that fall as the income increases, high earners benefit from tax allowances on private savings.}

One key element of the 1948 pension system---the different pension age for men and women---has been left unchanged until the last decade of the 20th century. Following the European Court rulings, the UK had to equalise male and female pension age, and in order to defuse the ``demographic bomb'', decided to gradually increase women's pension age to 65. Its subsequent increase to 68 for everyone in Pensions Act 2007 was motivated by rising life expectancies. The financial crisis spurred the Conservative-Liberal coalition government to accelerate this increase twice, in Pensions Acts 2011 and 2014.\footnote{In July 2017, the government decided to accelerate the state pension age rise to 68 again, following the recommendations of the Cridland report. It will now be phased in between 2037 and 2039, rather than from 2044 as was previously proposed.\cite{Cridland}}

As the mortality rates at all ages continue to fall thanks to improving standards of living and advances in medicine, senior citizens are the fastest growing segment of the UK population. In 1948, when the National Insurance Act was implemented, retiring men and women were expected to live additional 11.6 and 18.7 years, respectively. According to the latest Office for National Statistics projections, those numbers rose to 21.4 and 28.4~\cite{mortalityONS}. At the same time, fertility rates---after experiencing a series of demographic upheavals (collapsing during Spanish flu pandemic and World Wars, and rebounding in the 1920 birth-rate spike and during post-WW2 and the 1960s ``baby booms'', the latter reaching 2.93 children per woman at its peak in 1964)---fell and stabilised around the current average of 1.81 children per woman~\cite{birthsONS}. The longevity together with high immigration (mainly from the EU after the 2004 enlargement~\cite{migrationONS}) contribute to the growth of population size, despite the low birth rates~\cite{Werpachowska}.

The above factors shape the population age structure, and thus are important considerations for the provision of state pensions. Additionally, economic and demographic effects of Brexit (especially the free movement arrangements) will potentially have a significant impact on the affordability of the pension system~\cite{DEEUreport}.

\section{Mortality rates forecast using neural networks}
\label{sec:deeplearning}

Future mortality rates for all age and sex groups are required as inputs for our microsimulation. Rates of survival were consistently improving throughout the UK population for more than 100 years, owing to medical progress and bettering standards of living. A reasonable projection should continue this trend. How much further improvement in the lifespans of future generations can be expected is, however, far from certain, with some forecasts predicting over 30\% chance for a person born today of reaching the ranks of centenarians~\cite{ONSChancesOfHundred}. Inevitably, the further away in the future we project mortality rates, the higher is their uncertainty. Simple analytic models~\cite{LeeCarter,Cairns} provide more stability, but may not be flexible enough to reflect all trends in the data.


We propose a robust forecasting model based on the deep learning approach\,\cite{Goodfellow}. It uses a recurrent neural network, which is dedicated to the analysis of dynamic temporal behaviours, to exploit the information about mortality trends available from historical data~\cite{qxONS}. Given the last $N=40$ mortality rates for a fixed age group, our model uses a recurrent neural network to predict the $(N+1)$-th one and then feed it back as part of the input used to predict the $(N+2)$-th rate, etc. Differently than in~\cite{Hainaut}, we use a neural network to predict the mortality rates directly, without additional assumptions about the random walk model used to evolve latent parameters in time. The complete mathematical description of our model and the procedure of optimising $N$ is provided in the Appendix.

\begin{figure}[ht]%
\centering
\includegraphics[width=1\linewidth, trim=0px 10px 0px 0px]{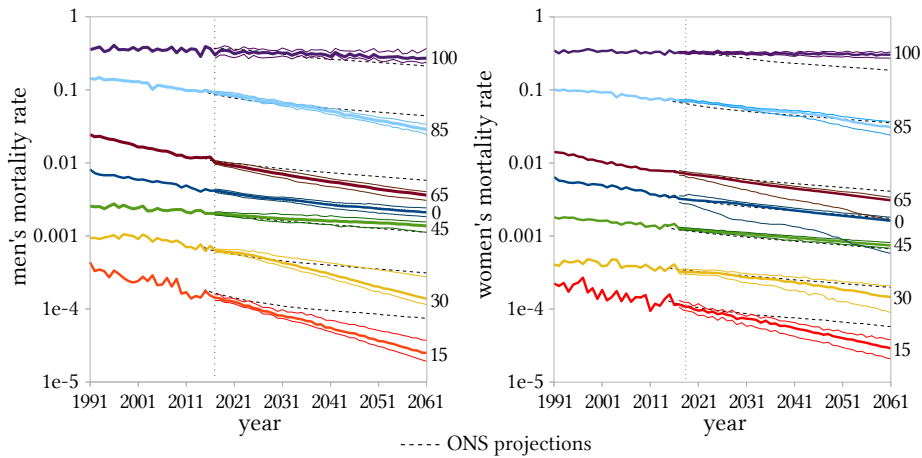}
\caption{Mortality rate forecasts (beyond 2016) obtained from the neural network model (solid lines) for men and women by age with 95\% bootstrap confidence intervals. Dashed lines represent projections from the ONS model~\cite{ONSmortalitymodel}.}%
\label{fig:mortality}%
\end{figure}

The obtained mortality rate forecasts are shown in Fig.\,\ref{fig:mortality} for selected age groups of men and women. The model training and extrapolation procedure has been repeated 100 times in order to remove the noise arising from the random initialisation of neural network parameters (see the Appendix). The mortality rates are calculated as medians of values from individual runs, while the 95\% confidence intervals are their 2.5\% (lower bound) and 97.5\% (upper bound) percentiles. The results indicate a progressing decline of mortality for men and women at all age. The men's mortality rates fall faster than women's. This trend is particularly strong for male teenagers, young adults, as well as ages 65--85, where it significantly diverges from the ONS results. Conversely, mortality rates for centenarians and senior women (above 85) do not fall as dramatically and remain at substantially higher levels than the ONS projections.

\begin{figure}%
\centering
\includegraphics[width=1\linewidth, trim=0px 18px 0px 12px]{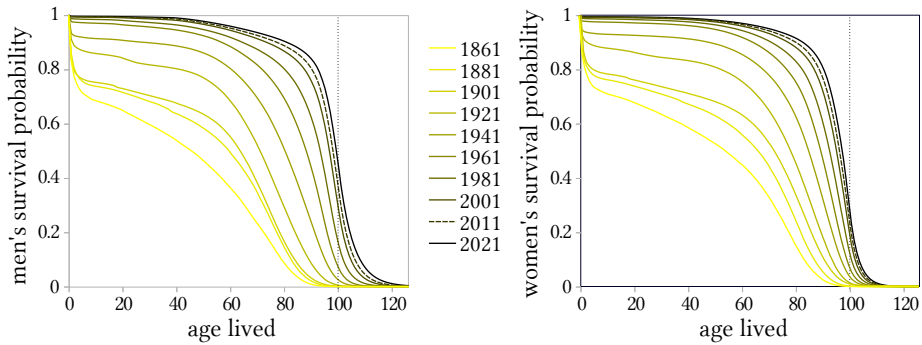}%
\caption{Survival curves for men and women in left and right panels, respectively. Rectangularisation can be observed for successive birth cohorts, particularly stronger for women.}%
\label{fig:survivalcurves}%
\end{figure}

The postponement of death to senectitude due to population ageing occurs with the rectangularisation of survival curves presented in Fig.\,\ref{fig:survivalcurves}. Furthermore, the age distribution in deaths is more compressed for women than men, giving the latter a noticeably higher chance of becoming centenarians, but also a higher risk of dying younger. An akin effect of men outliving women has been already observed in the British population. Men have been catching women up in average lifespan over the last decades, the first benefiting from the industry move from physical labour to services and adapting healthier lifestyle, while the second often taking the toll of combining full-time jobs with housework~\cite{life-expectancy-gap-closing}. At the same time, the positive correlation between life expectancy and socioeconomic status~\cite{men-outliving-women} combined with gender imbalance in the latter~\cite{socioeconomic-grade-by-sex} may cause an uplift in the men's age of death and the observed widening of its distribution.


\begin{figure}%
\centering
\includegraphics[width=0.7\linewidth, trim=0px 18px 0px 12px]{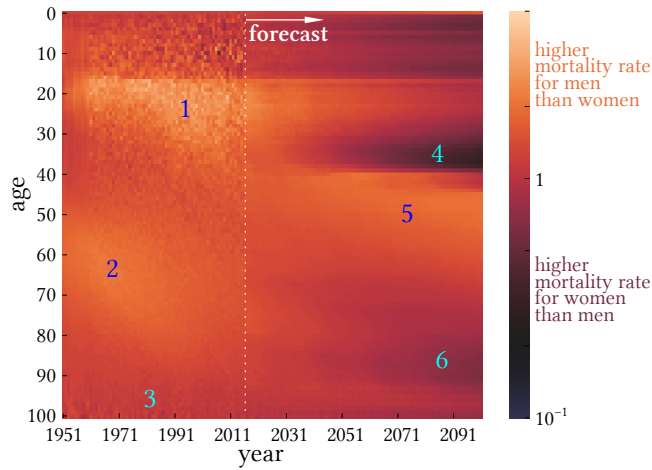}%
\caption{The ratio of contemporaneous mortality rates for men and women of the same age. Light regions (1, 2 and 5) indicate higher mortality rates for women than men, and dark ones (3, 4 and 6) otherwise.}%
\label{fig:mortalityratio}%
\end{figure}

Detailed historical and forecast patterns of relative mortality rates of men to women of the same age are shown in Fig.\,\ref{fig:mortalityratio} (in logarithmic scale). The bright regions indicate higher rates for women, while the dark ones higher rates for men. According to historical data, regions 1 and 2 are associated with four times higher prevalence of deaths from suicides, transport accidents and misuse of alcohol and drugs in young men, and twice higher prevalence of cardiovascular diseases in older men, as compared to their female peers, respectively~\cite{causes-of-death}. Region 3 corresponds to an increased number of women's deaths at old age, mainly due to the Alzheimer disease and dementia. The forecast reveals mortality trends transforming those patterns. In region 4, women in their thirties have higher mortality rates than men (although very low for both sexes), possibly due to the commonly reported problem of young women adapting unhealthy behaviours (smoking and drinking). The  weaker compression of mortality in men than women accounts for regions 5 and 6, which represent the lower and upper tails of men's distribution of age in deaths observed in the survival curves.

Since fairness between generations requires that everybody spends a similar proportion of adult life contributing to and receiving a state pension, it is interesting to compare the residual lifespan for people retiring under different legislations. Table\,\ref{tab:life_expectancy_retirement} displays life expectancies for cohorts who withdraw from employment in different years and at different ages: at the age of 60 for women and 65 for men, at the equal age of 65 and next 68. The life expectancies forecast by our model grow faster than the state pension age rises. In particular, women retiring in 2018 can expect to live 23 more years, which is almost 5 years longer than women retiring in 1948, despite their state pension age rising from 60 to 65. The difference is particularly striking for men (as it is not attenuated by the state pension age change), whose expected lifespan in retirement doubles. The following state pension age rise to 68 does not reduce the residual lifespan of those retiring in 2048, when it comes into effect (men gain another 1.5 year of life). While under the old legislation women and men could expect to spend 24\% and 15\% of their lives in retirement, respectively, the subsequent reforms guarantee a generous and fair approx.~26\% to next generations. Furthermore, they can expect to enjoy more years in good health~\cite{PHEhealthylifeexpectancy}, as the general health improvements that lead to increasing life expectancy can also delay the onset and progress of diseases, perhaps replacing them with natural death upon reaching a hypothesised biological limit of human lifespan~\cite{Fries}. This compression of morbidity scenario forms a more optimistic projection of the impact of population ageing on healthcare costs.


\begin{table}[ht]
\centering
\begin{tabular}{l|c|c|c}
 & 60 for women / 65 for men & Equal age (65) & Current reform (68)\\
\hline
1948 & 18.7 / 11.6 & 14.5 / 11.6 & 12.3 / 9.8\\
2018 & 28.5 / 22.2 & 23.4 / 22.2 & 20.7 / 19.3\\
2048 & 31.8 / 27.9 & 26.8 / 27.9 & 23.8 / 24.8
\end{tabular}
\vspace{2ex}
\caption{Life expectancy in retirement for cohorts of women and men retiring in different years and at ages indicated by different state pension systems reforms.}
\label{tab:life_expectancy_retirement}
\end{table}

\begin{figure}[ht]%
\centering
\includegraphics[width=0.96\linewidth, trim=0px 18px 0px 12px]{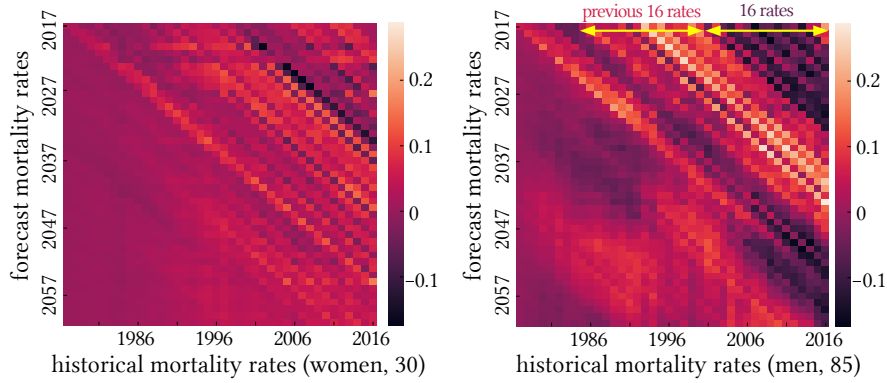}
\caption{Examples of sensitivities of the forecast mortality rates to historical data for two population groups.}%
\label{fig:sensitivity}%
\end{figure}

Figure\,\ref{fig:sensitivity} shows typical examples (for 30-year-old women and 85-year-old men) of sensitivities of forecast mortality rates to historical ones, which are calculated as ${\partial q_{t, i}}/{\partial q_{t', i}}$. The diagonal bands indicate a temporal autocorrelation in their trends with a length corresponding to the optimised input size $N=40$ (see Sec.\,\ref{sec:RNN-tuning}). It fades for longer time intervals, as expected from a recursively applied one-step extrapolation model with a fixed memory length. The obtained sensitivities are predominantly positive for young women, which is a hallmark of the general longevity growth in the population.

The results for old men reveal a more complex pattern, namely, the extrapolated mortality rate has first a broadly negative sensitivity to the preceding approximately 16 historical rates, and then a broadly positive sensitivity to the previous 16 historical rates. A simple mathematical reasoning\footnote{The Taylor series expansion of $q_{t,i}$ as a function of the preceding $N$ rates $q_{t',i}$ around a value $q_i$ representative of the group $i$ yields $q_{t,i} \approx \sum_{t'=t-N}^{t-1} \left. \partial q_{t,i}/\partial q_{t',i}\right|_{\boldsymbol{\xi}_i} (q_{t',i} - q_i) + c_i$, where $\boldsymbol{\xi}_i := [q_i, \dotsc, q_i]$ and $c_i := \exp{f(\boldsymbol{\xi}_i)}$. The diagonal stripes in Fig.\,\ref{fig:sensitivity} indicate that the sensitivities are translationally invariant in time, i.e.~$\partial q_{t,i}/\partial q_{t',i} \approx s_{t-t', i}$. Hence, $q_{t,i} \approx \sum_{t'=t-N}^{t-1} s_{t-t',i} q_{t',i} + c_i'$, where all terms independent of $t$ have been incorporated into $c_i'$. Since $s_{t-t'}$ is negative for $t-t' \le 16$, positive for $16 < t -t' \le 32$ and approaches zero for $t-t' > 32$, by further approximation we obtain $q_{t,i} - c_i' \sim \sum_{j=17}^{32} q_{t-j, i} - \sum_{j=1}^{16} q_{t-j, i}$. Therefore, an increasing long-term trend of $q_{t',i}$ tends to push $q_{t,i}$ down, while a decreasing one tends to push it up, a phenomenon often called \textit{reversion to the mean}.} indicates that the long term trend of mortality rate in this group has a tendency to ``revert to the mean'' (cf.~the mortality rate in Japan, which reverted in 1980s and grows consistently). As the life expectancy projections for Britain has been reported to flatline and slowly fall over the last years~\cite{life-expectancy-fall}, our result suggests that the falling trend may continue for a longer period of time owing to the long-term rise of mortality rate in older population predicted by our model.

In the next section we use the obtained mortality rates as an input to the microsimulation of the E\&W population.

\section{Pension cost dependency ratio estimation using microsimulation}
\label{sec:microsimulation}

Apart from the mortality rates forecasts obtained in the previous section, the microsimulation model employed in this work uses the same methods and data (updated with latest releases) as described in detail in Ref.\,\cite{Werpachowska}. It serves us to evaluate the impact of state pension reforms and Brexit by calculating the pension cost dependency ratio under two combined sets of scenarios.

\begin{figure}%
\centering
\includegraphics[width=0.59\linewidth, trim=0 18px 0 12px]{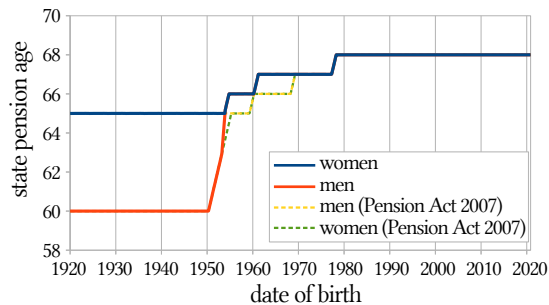}%
\caption{Timeline of changes to the state pension eligibility age by man or woman's birth date starting from 1921. State pension age established by consecutive reforms: National Insurance Act 1946 (implemented in 1948), European Court ruling 1995 equalising pension age to 65 and Pensions Act 2007 increasing it to 68, amended by Pensions Acts 2011 and 2014 accelerating the increase~\cite{timetables}.}%
\label{fig:spa}%
\end{figure}

The pension system scenarios assume different state pension age (SPA) with and without changes introduced by the past reforms as follows (Fig.\,\ref{fig:spa} displays their timeline):
\begin{itemize}[leftmargin=0mm,topsep=0.8mm]
\setlength\itemsep{0mm}
	\item[]\textbf{- Pre-reform}: SPA equal to 65 for men and 60 for women (as introduced by the National Insurance Act in 1948 and ignoring the subsequent reforms) over the whole simulation period.
\item[]\textbf{- Equal SPA}: SPA for women starts to increase in April 2010 under the Pensions Act 1995 by one month every month (by date of birth) to reach equal SPA of 65 for man and women in March 2020. The transition is accelerated in April 2016 by the Pensions Act 2011 to complete in November 2018. SPA equals 65 for everyone over the remaining simulation time.
\item[]\textbf{- SPA 68}: SPA for men and women will equalise in 2020 following the (overridden) Pensions Act 1995 and next rise to 66 between 2024 and 2026 (one month every month), to 67 between 2034 and 2036 and to 68 between 2044 and 2046 (Pensions Act 2007).
\item[]\textbf{- Accelerated SPA 68}: (current) Following the current legislation, SPAs for men and women equalise faster, i.e.\,in November 2018. Next, they rise to 66 between March 2019 and September 2020 (three months in the first month followed by one month every one of next nine months), to 67 between 2026 and 2028 and to 68 between 2044 and 2046 (Pensions Acts 2007, 2011 and 2014). 
\end{itemize}

The post-Brexit scenarios focus on migration patterns between the UK and the EU after March 2019, namely:
\begin{itemize}[leftmargin=0mm,topsep=0.8mm]
\setlength\itemsep{0mm}
\item[]\textbf{- Status quo}\ extrapolates current migration trends.
\item[]\textbf{- Soft Brexit}\ assumes amicable parting between the UK and the EU, followed by only small changes in migration flows of the UK and EU citizens. Just 10\% of the EU immigrants living in the UK will return to their country of origin (those who will have arrived closer to the Brexit date will leave first). This ``exodus'' will last 2 years and not involve the immigrants arriving after Brexit. Simultaneously, a wave of returns of British emigrants living in other EU countries will take place. We assume the repatriation of 10\% of their population. The level of regular migration flows of EU citizens to and from the UK will fall to the average of 2000 and 2011 figures, whereas the outflow of British citizens to the EU will be reduced by 80\%.
\item[]\textbf{- Hard Brexit}\ drastically limits the migration between the UK and the EU as many migrants lose the right of residence or decide to return to their country of origin. We assume the ``exodus'' of 70\% EU immigrants currently living in the UK and the repatriation of 80\% of the British living in the EU. The migration of EU citizens to the UK will return to much lower levels from before the 2004 EU enlargement, while the outflow of Britons to the EU will decrease by 30\%.
\end{itemize}

\begin{figure}%
\centering
\includegraphics[width=0.86\linewidth, trim=0 18px 0 12px]{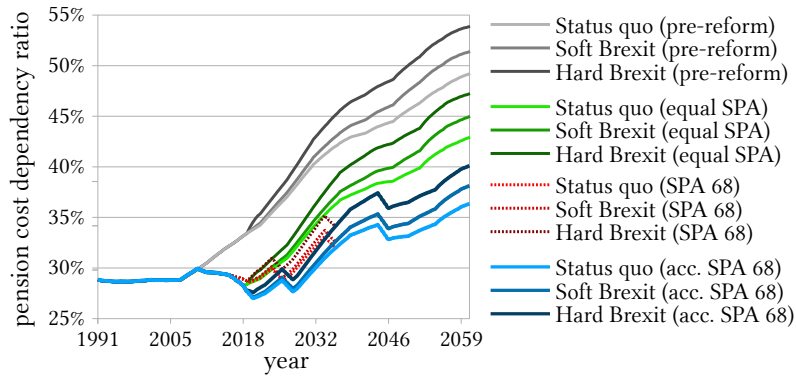}%
\caption{Pension cost dependency ratio under different state pension reforms and different post-Brexit demographic scenarios (as described in Sec.\,\ref{sec:microsimulation}).}%
\label{fig:dependencyratio}%
\end{figure}

Figure\,\ref{fig:dependencyratio} shows the trends of pension cost dependency ratio under the above scenarios. The historical values of the ratio maintained a sustainable level until 2007. Beyond that period it began to rise owing to the population ageing. The unprecedented rate of this process was propelled by post-war and 1960s baby boomers entering the state pension age. In addition, their retirement considerably reduces the workforce, whose large part consists of cohorts born in the period of declining fertility rates, from the mid 1960s to 1970s. The trend is somewhat softened by the influx of EU workers, especially the post-2004 enlargement wave.

The rise of state pension age for women starting in 2010 reverses the trend sharply, creating a substantial gap between the old and the new pension scheme scenarios. In particular, the reform reduces the pension cost dependency ratio to the level from 2007. Afterwards, the ratio picks up its previous upward trend fuelled by the advancing population ageing, with temporary reversions due to consecutive rises of retirement age. The rapid growth from 2027 to 2038 is caused by retiring 1960s ``baby boomers''. The growth subsides as the forecast life expectancies stop to grow, and takes off again around 2049, when the children of large 1960s cohorts enter their retirement age.

The magnitude and trend of pension cost dependency ratio strongly depend on the considered state pension reforms and are moderately affected by the varying migration patterns in different EU membership scenarios. In particular, the currently introduced state pension age reforms will lead to a considerable and permanent reduction of the ratio, almost reverting it back to the levels from before 2007. However, in the long term---in particular after the large 1960s cohorts achieve the retirement age---the ratio will grow again under all considered scenarios owing to the population ageing.

\section{Summary}

We have used our stochastic dynamic microsimulation model~\cite{Werpachowska} to calculate detailed forecasts of the pension cost dependency ratio for England and Wales until 2061, with and without taking into account the series of retirement age rises under the state pension system reforms initiated by the Labour and continued by the Conservative governments as well as considering changes in international migration patterns under different Brexit scenarios. To fully account for the changing trends in life expectancies, we have implemented a custom extrapolation model for mortality rates, based on deep learning techniques. It has discovered complex patterns in historical and forecast mortality statistics: the survival curve rectangularisation with a stronger compression of the age of death distribution for women, an arising trend of men outliving women and a possible sustained reversion of mortality rate trend from decrease to growth. Our results show that the recent reforms, although politically controversial, can be expected to stave off the ``pension crisis'' and bring back the system on a sounder fiscal footing in the following decade. At the same time, more workers can expect to spend longer share of their lifespan in retirement, despite the state pension age rises. If the same factors promoting longevity also delay the disease processes in the lifespan, they can expect to live more years in good health, without putting an additional strain on the healthcare system. To a lesser degree, the future pension cost dependency ratio will depend also on the post-Brexit relations between the UK and the EU, with ``softer'' alignment lowering the relative cost of the pension system (due to the expected continued influx of young workers from the EU) and ``harder'' one raising it. In the long term, however, a further increase of the state pension age may be required if the current population trends persist.

\appendix
\section{Mortality rate extrapolation model}
\label{sec:mort-rate-extrap-model}

\subsection{Data}

The $qx$ mortality rate $q_{t,i}$ is the probability that a person of age $i$ (measured in years) alive at time $t$ will die within the next year. The vector of historical mortality rates for all ages $i=0, \dotsc, M$ is denoted by $\vec{q}_t$, where $t = T_i, \dotsc, T_f$ is time in years. In our case $M = 100$, $T_i = 1951$ and $T_f = 2016$. We analyse data on the E\&W population by age and sex (men or women).

\subsection{Extrapolation model}

It can be noted that for many age groups $i$ the logarithm of the mortality rate $q_{t,i}$ is approximately linear in $t$ (see Fig.\,\ref{fig:mortality}). Accordingly, we apply the extrapolation model to log-rates $x_{t,i} := \ln q_{t,i}$. It has the additional benefit of imposing the constraint $q_{t,i} = e^{x_{t,i}} \ge 0$.

We propose a recurrent neural network~\cite{Goodfellow} extrapolation model which, given a vector $\vec{\xi}_{t-1,i} := [ x_{t - N, i}, \dotsc, x_{t-1, i} ]$ of $N$ consecutive log-rates, 
outputs the value of $x_{t,i} = f(\vec{\xi}_{t-1,i})$. To extrapolate beyond time $T_f$, the model is applied recursively to $\vec{\xi}_{T_f, i}$ for each $i$. We denote the result of applying $f$ recursively $k$ times by $f^{(k)}$, with $f^{(1)} \equiv f$. Thus, the extrapolated log-rate $x_{T_f + k, i} := f^{(k)}(\vec{\xi}_{T_f, i})$. For example,
\begin{equation}
x_{T_f+2,i} = f^{(2)} (\vec{\xi}_{T_f, i}) = f([x_{T_f-N+2, i}, \dotsc, x_{T_f,i}, f^{(1)}(\vec{\xi}_{T_f,i})])\ .
\label{eq:recursive}
\end{equation}
The function $f: \mathbb{R}^{N} \to \mathbb{R}$ is calculated by the \emph{cell} of the network: a composition of linear and non-linear mappings between the input $\vec{\xi}_{t-1,i}$ and output $x_{t,i}$.

The cell used by our model is a fully-connected neural network consisting of $K$ neuron layers $\vec{z}_k \in \mathbb{R}^{D_k}$, $k=0,\dotsc,K-1$. The \emph{input layer} $\vec{z}_0$ is equal to the argument of function $f$, $\mathbb{R}^N \ni \vec{z}_0 = \vec{\xi}_{t-1,i}$ and $D_0 = N$. Every subsequent layer is a function of the previous one, of the following form
$\vec{z}_k = h_k(\mat{W}_k \cdot \vec{z}_{k-1} + b_k)$,
where $h_k: \mathbb{R} \to \mathbb{R}$ is the $k$-th \emph{activation function}, $\mat{W}_k \in \mathbb{R}^{D_k \times D_{k-1}}$ is the $k$-th \emph{weight matrix} and $\vec{b}_k \in \mathbb{R}^{D_k}$ is the $k$-th \emph{bias vector}. For $k < K-1$, $h_k$ is the so-called ReLU (\emph{rectified linear unit}) function, while $h_{K-1}$ is the identity:
\[
h_k(z) = \begin{cases}
\max(0, z) & k < K - 1 \\
z & k = K - 1
\end{cases} \ .
\]
The intermediate layers $0 < k < K-1$ are called \emph{hidden layers} and have dimension $H$. The last, \emph{output layer} is the extrapolated value, $\mathbb{R}^1 \ni \vec{z}_{K-1} = [x_{t+1, i}]$, with $D_{K-1} = 1$.

\subsection{Model initialisation and training}

Before training, bias vectors and weight matrices need to be assigned initial values. In our model, all bias vector elements are initialised to constant value $b_\text{init} > 0$ and all weight matrix elements are drawn independently from $N(0, \sigma_\text{init})$ distribution truncated at two standard deviations (larger values were discarded and re-drawn). Such initialisation helps to avoid saturating the nonlinear activation function and breaks the symmetry between different units~\cite{Goodfellow}. 

Training the network consists in minimising the training loss $c[f]$, namely, the average $\ell^2$ error between \emph{training targets} $\vec{y}_{t,i} := [x_{t, i}, \dotsc, x_{t + {N_\text{train}} - 1, i}]$ and \emph{training outputs} $\hat{\vec{y}}_{t,i}  := [f(\vec{\xi}_{t-1, i}), f^{(2)}(\vec{\xi}_{t-1, i}), \dotsc, f^{(N_\text{train})}(\vec{\xi}_{t-1, i})]$ 
(see~Eq.\,\ref{eq:recursive}):
\[
c[f] := \frac{ \sum_{i=0}^M \sum_{t=T_i + N}^{T_f - N_\text{train} - 1} \lVert \vec{y}_{t,i} - \hat{\vec{y}}_{t,i} \rVert_2^2 }{(M+1) (T_f - T_i + 1 - N - N_\text{train} + 1)}\ .
\]
The loss $c[f]$ is minimised over the neural network weights and biases using the RMSProp~\cite{RMSProp} variant of the Stochastic Gradient Descent method~\cite{Goodfellow}. A crucial aspect of this method is that $c[f]$ is not calculated over all pairs of $\vec{y}_{t,i}$ and $\hat{\vec{y}}_{t,i}$, as in the formula above, but over a randomly sampled \emph{mini-batch}~\cite{Goodfellow,RMSProp} of size $B$. The training is run for $L_\text{train}$ steps of the RMSProp optimiser. The learning rate of the optimiser is set to initial value $\alpha_0$ and then multiplied by a factor $0 < \kappa < 1$ every $\Delta L$ steps if the average training loss over the last $\Delta L$ steps did not decrease compared to the average over previous $\Delta L$ steps.

\subsection{Model configuration}
\label{sec:model-config}

The model parameters are set as follows: input size $N=40$, number of layers $K=6$ and number of hidden neurons per layer $H=64$. The three parameters and the number of training steps $L_\text{train}=270\,000$ were tuned using a validation set, as described in the next section. The remaining training parameters, i.e.\,training sequence length $N_\text{train}=10$ (relatively small in order to generate more training sequences), mini-batch size $B=16$, learning rate adjustment interval $\Delta L=10\,000$, initial learning rate $\alpha_0=10^{-4}$ and learning rate adjustment factor $\kappa=0.9$, as well as initialisation parameters, i.e.\,initial bias $b_\text{init}=0.1$ and truncated standard deviation of initial weights $\sigma_\text{init}=0.1$, were set to reasonable values.

\subsection{Implementation, parameter tuning and testing}
\label{sec:RNN-tuning}

The numerical code~\cite{github} has been implemented in Python and TensorFlow~\cite{TensorFlow}.

To determine the optimal values of model parameters and the number of training steps, we split the historical data (for each sex separately) into three sets: \emph{validation set} (containing age groups 1, 6, 11, $\dotsc$, 91 and 96), \emph{test set} (containing age groups 2, 7, 12, $\dotsc$, 92 and 97) and \emph{training set} (containing all remaining age groups). We then train the model on the training set for 300\,000 steps, testing several combinations of different values of $N$ (15, 25 and 40), $K$ (3--7) and $H$ (32, 64, 128, 256 and 512). Each configuration is tested for data for men and women separately. Every $\Delta L = 10\,000$ steps we calculate the average loss on the \emph{validation set} and record the lowest achieved validation loss during the entire training period, as well as the validation loss after the maximum 300\,000 training steps (the procedure is guided by the observation that, during training, the validation set loss at first decreases along with the training set loss, but after some time it starts to increase, indicating the onset of overfitting). The final and minimum validation set loss is found to be most affected by $N$, while $K$ and $H$ compensate for each other (the higher $K$, the lower $H$, and vice versa). Based on the validation set loss values, we choose the best values of $N$, $H$, $K$ and $L_\text{train}$, which are common for both sexes, as reported in Sec.\,\ref{sec:model-config}.


To confirm the model performance using the parameters selected above, we retrain it (separately for men and women) on the combination of training and validation sets, measuring the final loss and final bias (average difference between predicted and target log-rates) on the \emph{test set}. 
The results for men and women, respectively, are as follows: final training and validation set loss is 0.00682 and 0.00696, final test set loss is 0.0106 and 0.0122 and final test set bias is 0.0241 and -0.0255.

\renewcommand{\refname}{References}


\end{document}